\documentclass[apj,twocolumn]{openjournal}
\usepackage{amsmath}
\usepackage{booktabs}
\usepackage{multirow}
\usepackage{color}
\usepackage{soul}
\usepackage[dvipsnames]{xcolor} 
\definecolor{ultra}{HTML}{0B6572}
\definecolor{udark}{HTML}{0A5681}
\usepackage[breaklinks,colorlinks,citecolor=blue,urlcolor=blue,linkcolor=blue]{hyperref}
\renewcommand{\arraystretch}{1.2}  
\usepackage{listings}
\usepackage{orcidlink}
\allowdisplaybreaks

\begin{document}


\title{Relationship Between Major Stellar Physical Parameters and Normal Mode Frequencies in Accreting White Dwarf Stars}

\author{\vspace{-40pt}Praphull Kumar\orcidlink{0000-0002-8791-3704}$^{1}$}
\author{Dean M. Townsley\orcidlink{0000-0002-9538-5948}$^{1}$}
\author{Hunter Anz$^{1}$}

\affiliation{$^1$Department of Physics \& Astronomy, The University of Alabama, Tuscaloosa, Alabama, USA}

\email{Corresponding author: pkumar5@crimson.ua.edu}
\email{praphullkr05@gmail.com}

\title{Relationship Between Major Stellar Physical Parameters and Normal Mode Frequencies in Accreting White Dwarf Stars}

\begin{abstract}

White dwarfs (WDs) are the final fate of about 97\% of the stars in our galaxy, making them vital tracers of stellar history. A fraction of WDs exist in cataclysmic variable (CV) systems, accreting matter from a nearby companion star. A subset of CVs undergo episodic rapid mass transfer, termed dwarf novae (DNe) outbursts. Some accreting WDs exhibit near sinusoidal photometric variations, interpreted as $g$-mode pulsations. However, identifying pulsation modes in accreting WDs remains challenging due to the paucity of available observed modes. In this work, we present a comprehensive computation of the observable $g$-mode frequencies across a range of WD parameters, varying the WD mass, size of the newly accreted layer and core temperature. We also introduce a novel method for mode identification based on the time evolution of pulsation periods following an accretion episode. Our mode identification method does not rely on the direct detection of the consecutive radial mode orders, frequently required in isolated WDs. Moreover, this work improves upon our previous WD modeling efforts. We use a more realistic core temperature in addition to thermohaline mixing and element diffusion enabled during the accretion phase.

\end{abstract}


\section{Introduction} \label{sec:intro}
A typical cataclysmic variable (CV) contains a primary white dwarf (WD) star accreting material from a less massive main sequence companion star via Roche lobe overflow. During the longest-lived evolutionary phase, the accretion rate is low, averaging $\langle\dot{M}\rangle\sim 10^{-11}-10^{-10}~M_\odot ~\rm yr^{-1}$ \citep{Howell_2001}. Prolonged accretion ($\sim 10^7$ -- $10^8$ years) onto a WD primary leads to an unstable thermonuclear runaway, a hydrogen flash, on its surface, resulting in a classical nova (CN). In many CV systems, particularly those with thermally unstable accretion disks, the WD undergoes rapid mass transfer rates of $\langle\dot{M}\rangle\geq 10^{-9}~M_\odot ~\rm yr^{-1}$ for an active time of several days up to a month or so, recurring on months to years timescales. These accretion events are observed as dwarf novae (DN) outbursts \citep{Warner_1995}. During the quiescence window between outbursts, the mass transfer rate is small enough that the system's UV and sometimes optical light is dominated by photospheric emission from the WD, allowing the measurement of its effective temperature, $T_{\rm eff}$ \citep{Sion_1999, Pala_2017}. While a comprehensive understanding of CV evolutionary behavior remains unclear, particularly regarding how the typical WD mass observed comes to be and how the angular momentum of the WD evolves over time \citep{Hillman_2016, Schreiber_Zotorovic_Wijnen_2016}, research continues to advance our knowledge of these fascinating objects.

Since the first detection of non-radial pulsations in an accreting white dwarf (GW Lib; \cite{Warner_1998, Vanzyl_2004}), the number of such systems exhibiting pulsations has significantly increased \citep{Szkody_2002, Szkody_2004, Szkody_2005, Gansicke_2006, Mukadam_2013, Szkody_2021}. Asteroseismology holds the potential to provide further insights into the internal structure of the white dwarf and the ongoing evolution of the system that contains it. The pulsations in accreting WDs are conceptually similar to those observed in isolated counterparts, primarily characterized by gravity modes ($g$ modes) \citep{Brassard_1991, Fontaine_2008, Romero_2012}. These $g$-mode oscillations are restored by the buoyancy force. Recently, \cite{Kumar_Townsley_2026} continued to argue that pulsations on accreting WDs are best explained by $g$ modes that are thermally driven by the outer convection zone. The related Rossby mode ($r$ mode) oscillations are disfavored based on what is known about the driving and the observed frequencies. Accreting WDs have atmospheres composed of H and He similar to the companion main sequence star. The $g$ modes are driven by their thermal interaction with the convection zone that arises from the ionization of H and He near the surface of the WD, similar to the mechanism responsible for DAV and DBV stars \citep{Brickhill_1991, Wu_Goldreich_1998, Arras_Townsley_Bildsten_2006, Van_Grootel_2015}.

While pulsations in isolated WDs like ZZ Ceti and V777 Her have been extensively studied \citep{Winget_and_Kepler_2008,Fontaine_2008, Aerts_2010, Althaus_2010, Romero_2012}, our understanding of pulsations in rapidly rotating accreting WDs remains limited \citep{Townsley_Arras_Bildsten_2004, Saio_2019, Kumar_Townsley_2023, Kumar_Townsley_2026}. In this work, we aim to continue to lay the groundwork for interpreting the oscillations seen in accreting WDs through a forward modeling approach. WD asteroseismology holds a great potential to reveal intrinsic stellar properties by constraining the thickness of the newly accreted mixed layer, stellar mass, properties of the core composition layers, and more. 

As the WD cools after an accretion event, each mode frequency relaxes to its pre-outburst value over a few months, as computed by \cite{Kumar_Townsley_2023} (see their figure 11). The unique rate of change for each mode may make it possible to identify the order of individual modes without having to measure a large number of modes. From an observational perspective, this is dependent on monitoring the WD's photometric variations for an extended period following the an accretion heating event. In this work, along with our variation of properties of the WD, we explore the use of the time derivative of mode periods, $\dot P$, as an indicator of mode identity. 

We continue to progress toward a more realistic WD model that incorporates more physics and better characterization of evolution. Compared to \cite{Kumar_Townsley_2023}, we now include element diffusion during the accretion process, rather than only during the WD cooling. This is crucial for establishing a physically valid structure for the core-shell boundary, which influences the mode frequency spacings. We have adopted an accretion rate more consistent with that anticipated for the stage of binary evolution that CV WD systems are in \citep{Townsley_and_Gansicke_2009}, necessitating a lower core temperature to match observed effective temperatures \citep{Townsley_Bildsten_2004}. Challenges remain in computing H shell flashes due to the essential role of convective boundary mixing. 

The purpose of this work is to explore how mode frequencies vary with changes in the major structural features of an accreting WD. A description of the parameter space explored and the evolutionary scenario and assumptions used to create a representative accreting WD structure and seismology are discussed in section \ref{sec:method}. The results of our mode frequency calculations are shown in section \ref{sec:results}. We close this paper with discussion and conclusions in section \ref{sec:discussions}.
\section{Method}\label{sec:method}

In this section, we describe the making of the WD models and elaborate on the choice of parameters used in this study. Sections \ref{ssec:wdmodelsbeforeaccretion} to \ref{ssec:dwarfnovae} extend the discussion of the evolutionary stages of the WD, along with highlighting differences from our previous works \citep{Kumar_Townsley_2023, Kumar_Townsley_2026}. Section \ref{ssec:choiceofparameter} outlines the choice of parameter space used for the WD models and how it relates to what is known for the CV WDs. Section \ref{ssec:choiceofmodes} outlines our choice of modes that we expect might be seen prominently on accreting WDs.

\subsection{WD Models Before Accretion}\label{ssec:wdmodelsbeforeaccretion}
Here we describe the detailed modeling of the CV WDs and the models used to compute the normal modes. We use the one-dimensional stellar evolution code Modules for Experiments in Stellar Astrophysics \citep[\texttt{MESA;}][]{Paxton_2011, Paxton_2013, Paxton_2015, Paxton_2018, Paxton_2019}, to perform our stellar evolutionary calculations. The general layout of constructing the WD is similar to what is used in \cite{Timmes_2018, Kumar_Townsley_2023, Kumar_Townsley_2026}. In total, we consider five different WD mass models. Two of these are adopted from \cite{Kumar_Townsley_2023}, with their final masses of 0.78~$M_\odot$ (initial mass = 3.95~$M_\odot$) and 0.93~$M_\odot$ (initial mass = 6.1~$M_\odot$). While we used their models after the Asymptotic Giant Phase (AGB) as initial configurations, we cooled them using the more extended \texttt{pp\_cno\_extras\_o18\_ne22.net} reaction network including the element diffusion, rather than the \texttt{basic\_network.net} employed in their work. In addition, all models in this study were computed with \texttt{MESA} release version r15140, which provides improved timestep control compared to the version r10398 used by \cite{Kumar_Townsley_2023}. 

For our 0.86~$M_\odot$, we continue to use \texttt{MESA} version r15140. We evolved a model starting from the pre-main sequence with an initial mass of 5~$M_\odot$ and solar metallicity ($\mathrm{Z=0.02}$), utilizing the 49-isotope reaction network. However, due to the well-known numerical difficulties near the AGB phase, we stall the simulation when the WD envelope mass reaches below $1.62\times 10^{-4}~ M_\odot$. We then stripped off the remaining hydrogen from the surface, yielding a hot WD that was subsequently cooled using a slightly smaller \texttt{pp\_cno\_extras\_o18\_ne22.net} reaction network (reduces the computational cost), incorporating element diffusion until the $T_{\rm eff}$ reaches 15,000 K.  

The remaining two WD models, with masses of 0.60~$M_\odot$ and 0.70~$M_\odot$, are rescaled from 0.78~$M_\odot$ using the \texttt{relax\_mass\_scale} method in \texttt{MESA}. Before applying the rescaling, the post-AGB hot 0.78~$M_\odot$ model from \cite{Kumar_Townsley_2023} is cooled for another $\sim 10^6$ years while we transition to \texttt{MESA} version r15140 from r10398. While this relaxation method preserves the internal composition profiles, it does lead to what is initially a slightly unusual thermal profile. However, these are not expected to significantly influence the pulsation behavior, as normal mode calculations are performed after the model has been evolved further. Once the WD masses are relaxed to 0.60~$M_\odot$ and 0.70~$M_\odot$, these models are cooled using the  \texttt{pp\_cno\_extras\_o18\_ne22.net} reaction network including element diffusion, until their core temperatures ($T_{\rm c}$) reached $6\times10^6$~K \citep{Paxton_2015}. We continue to use this reaction network for all subsequent WD evolution.

\subsection{Long-Term Time Averaged Accretion}\label{ssec:longtermaccretion}
The freshly accreted hydrogen and helium in a mass-transferring binary system induce an unstable thermonuclear ignition on the surface, resulting in a Classical Nova (CN) outburst \citep{Townsley_Bildsten_2004, Nomoto_2007, Schreiber_Zotorovic_Wijnen_2016}.
The cooled WD is allowed to settle via further cooling for $\sim 10^5$ years without accretion or element diffusion. Accretion is then initiated, still without element diffusion, for another $\sim 10^5$ years (relatively short period compared to overall WD lifetime) and with convection turned off. This prevent any premixing between the freshly accreted material and the helium remnants on the WD's surface leftover from the previous evolutionary stage. The interface between the layers created by this process is not critically important anyway since it is thin enough to be smoothed out by later diffusion. Following this, the WD is subjected to long-term accretion with rate, $\dot M$, chosen to give the desired physical state as described in section \ref{ssec:choiceofparameter}. Element diffusion and thermohaline mixing are included throughout the accretion phase. We also enable predictive mixing during accretion, as it helps stabilize the intermittent mixing regions that result from the interplay of accretion, diffusion, and thermohaline. Unless otherwise required by parameter choices, we use a rate of $ \dot {M} = 6\times 10^{-11}~M_\odot$~yr$^{-1}$ or the rate appropriate to the relevant long-term $\dot M$ for that case. This approach is essential for ensuring a physically valid core-shell boundary structure, which directly influences mode frequency spacings. We perform this long-term accretion on all the five different WD mass models, along with additional lower $T_{\rm c}$ models of a 0.78~$M_\odot$ (see below for details on our parameter choices). Rotation is included from the beginning through to the cooling phase of the WD. However, we do not include rotation during the accretion phase, instead, we explicitly incorporate rotational effects in \texttt{GYRE} \citep{Townsend_2013, Townsend_2018} when computing the normal mode frequencies.

The freshly accreted layer, $M_{\rm acc}$, is parameterized by the mass. The accreted mass fractions are $X_{\rm H} = 0.742$, $X_\mathrm{{4_{He}}} = 0.237$, $X_{\rm Z} = 0.02$, and $X_\mathrm{{3_{He}}} = 0.003$. As described by \cite{Townsley_Bildsten_2004}, a trace of $^3{\rm He}$ from the donor \citep{Shara_1980} can lead to early ignition at low temperatures. We use the mixing length parameter $\alpha_{\mathrm{mlt}} = 1.5$, $\alpha_{\mathrm{semi-convection}} = 0.01$, and $\alpha_{\mathrm{thermohaline}} = 1$. Due to persistent numerical challenges in modeling multiple hydrogen shell flashes with convective boundary mixing, we instead focus on accretion onto a bare white dwarf and evaluate its seismological properties prior to the onset of the hydrogen shell flash. This alternative is preferable to simulating an unrealistic, numerically unstable, and unconstrained mixed layer. Therefore, we halt the accretion evolution when the hydrogen surface luminosity of the WD $\log L_{\rm H} \simeq$ 5 \textrm{$L_{\odot}$} (where $L_{\odot}$ is the solar luminosity), accumulating mass immediately prior to the explosion or the next CN cycle, $M_{\rm ign}$. The maximum envelope mass $M_{\rm acc}$ leading to $M_{\rm ign}$ is set by the unstable nuclear burning. This process is mostly governed by the proton-proton ($p-p$) chain, near the base of the accreted H/He layer that causes the CN runaway. As the accreted material builds up on the surface, the temperature and pressure at the base of the envelope increase, finally triggering the runaway, marking the ignition point with setting the upper limit on $M_{\rm acc}$ before the outburst. To make a note, we do not use the ignition model; it is only used to set the $M_{\rm ign}$, so that the models correspond to a comparable fraction of the path toward ignition.

\subsection{Short-Term Accretion or Dwarf Novae} \label{ssec:dwarfnovae}

Dwarf novae (DNe) are the subset of the CVs \citep{Warner_1995} that undergo dramatic, episodic accretion events onto the WD surface. These outbursts last from a few days to months, with a high accretion rate of $\sim 10^{-8}M_\odot$~yr$^{-1}$, followed by a quiescent state lasting from days to decades. During quiescence, the accretion rate onto the WD drops significantly, allowing a more direct inspection of the WD surface, most notably $T_{\rm eff}$. A far-ultraviolet flux spectroscopic examination of the underlying accreting WDs reveals that they cool to their pre-outburst quiescent temperatures \citep{Pringle_1988, Long_1993, Szkody_2012}, a phenomenon that has also been supported by theoretical models \citep{Piro_2005, Kumar_Townsley_2023}.

Following the long-term accretion phase, we transition to the modeling of the DN (dwarf nova) cycle. During this phase, the WD experiences intermittent accretion with a rate of $\dot{M}=$ 1.2~$\times$~10$^{-8} M_\odot$~yr$^{-1}$ for a short duration of two months, with a recurrence time of 30 years. We evolve through three DN cycles before utilizing the model for seismology. We particularly investigate the characteristics of $g$ modes (see below for the rationale behind this choice).
\begin{figure}[h]
    \centering
    \includegraphics[width=1.\linewidth]{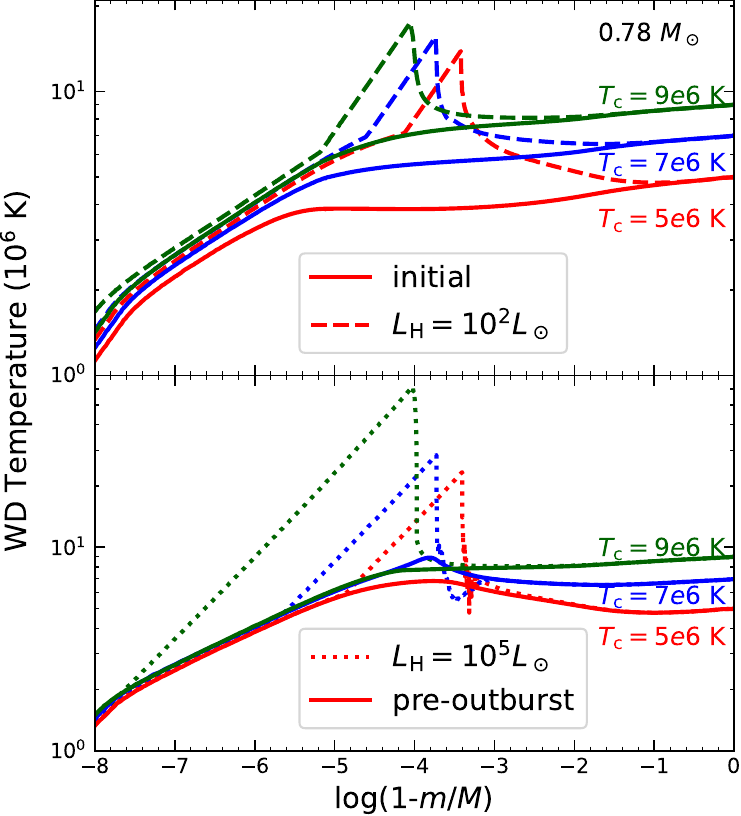}
    \caption{Temperature profiles against the fractional depth ($\log (1-m/M)$) at the beginning of the long-term accretion phase (top panel, solid) and the hydrogen luminosity $\log(L_{\rm H}/L_\odot)=2$ (top panel, dashed), pre-outburst (bottom panel, solid), and $\log(L_{\rm H}/L_\odot)=5$  (bottom panel, dotted), for three different core temperatures with $T_{\rm c} = 5,7$, and $9\times 10^6$~K of a 0.78~$M_\odot$ WD model.}
    \label{fig:TemperatureprofiledifferentTc}
\end{figure}

\begin{figure}[h]
  \includegraphics[width=1.\linewidth]{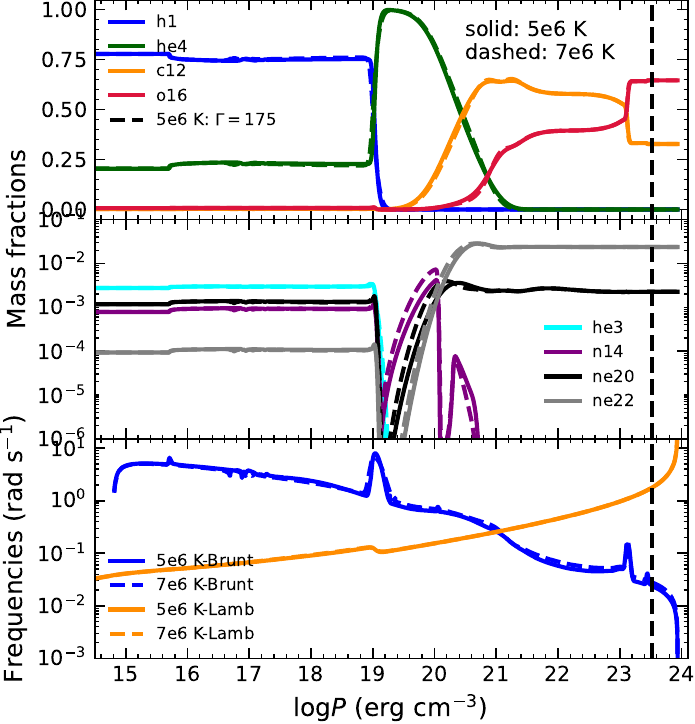}
   \caption{WD compositions and frequency profiles after a long-term accretion phase with $M_\text{acc} = 1.5\times 10^{-4}~M_\odot$ at core temperatures $T_{\rm c} = 5$ and 7~$\times 10^{6}$~K of a 0.78~$M_{\odot}$ WD model at $T_{\rm eff} = 14000$ K. Brunt-V\"ais\"al\"a and Lamb frequency are shown in the bottom panel. Mass fractions are shown on the log scale in the middle panel to highlight the other significant species produced that are difficult to note on the linear scale (top panel). The fraction of solid core is denoted by the vertical dashed line and is about 34\% in radius for $T_{\rm c} = 5\times 10^6$~K. The WD is uncrystallized at $T_{\rm c} = 7\times 10^6$~K.} 
   \label{fig:Abundanceandbrunt0.78M5e67e6K} 
\end{figure}

\subsection{Description of the Parameter Space}\label{ssec:choiceofparameter}

We now present a comprehensive description of the parameter space employed in this study, along with the rationale behind our choices. Thanks to the $Gaia$ mission \citep{Gaia_2016, Gaia_2018} and the ultraviolet data from the $Hubble ~ Space ~ Telescope$, the number of CVs observed with accurately determined WD masses, $M_{\rm WD}$ and $T_{\rm eff}$, has significantly increased, now exceeding over one hundred \citep{Pala_2017, Shara_2018, Pala_2022}. CV binaries are believed to lose their angular momentum through either gravitational radiation or magnetic breaking depending on their orbital periods \citep{Kolb_Baraffe_1999, Howell_2001, Townsley_Bildsten_2003}. This angular momentum loss determines the average accretion rate $\langle\dot{M}\rangle$ onto the WD. The WD's $T_{\rm eff}$ traces this process, as it reflects compressional heating from the accreted material \citep{Townsley_Bildsten_2004, Townsley_and_Gansicke_2009}, thereby constraining $\langle\dot{M}\rangle$ \citep{Townsley_Bildsten_2003}. A relatively good estimates of $\langle\dot{M}\rangle$ may inform the accumulated mass, $M_{\rm acc}$, and the ignition mass, $M_{\rm ign}$\citep{Townsley_Bildsten_2004}. Hence, we anchor our study on this crucial parameter, $T_{\rm eff}$. 

We break our seismological study into three independent scenarios with: (1) Models with diverse core temperatures, $T_{\rm c}$, while keeping the accreted layer mass, $M_{\rm acc}$, fixed for a 0.78~$M_\odot$ mass model, (2) varying $M_{\rm acc}$ at a fixed $T_{\rm c} = 5\times 10^{6}$~K for a 0.78~$M_\odot$ mass model, and (3) models with five different WD masses, $M_{\rm WD}$ at a fixed $T_{\rm c} = 6\times 10^{6}$~K. In each individual case, $\langle \dot{M}\rangle$ is adjusted to give a $T_{\rm eff}$ of 14000~K. In our previous work, \citep{Kumar_Townsley_2023}, we adopted a lower accretion rate ($\sim 10^{-12} M_\odot$~yr$^{-1}$) and correspondingly higher $T_{\rm c}$ ($\sim 10^7$~K). However, the recent results from \cite{Pala_2022} suggests that the WDs in CVs below the period gap accrete at rates higher than previously assumed. Based on their reported $\langle\dot{M}\rangle$ and $\langle\dot{M}\rangle$ derived from $T_{\rm eff}$ \citep{Townsley_Bildsten_2003}, WDs in CVs are expected to have cooler $T_c$ \citep{Townsley_Bildsten_2004}. Therefore, in this study, we adopt an accretion rate of $\sim 10^{-10} M_\odot$~yr$^{-1}$, consistent with expectations for CVs at their current evolutionary stage \citep{Townsley_and_Gansicke_2009}, correspondingly lower $T_{\rm c}$ in order to match observed $T_{\rm eff}$ \citep{Townsley_Bildsten_2004}. For a 0.78~$M_\odot$, the equilibrium core temperatures, $T_\text{c, eq}$, falls within a range of $5-7\times 10^6$~K.  

\subsubsection{Varying Central Temperature}\label{sssec:varycentraltemp}
The model parameters are summarized in Table \ref{table:table_all_parameters}.
It is important to note that we do not actively investigate the $T_\text{c, eq}$ for a given value of $\langle\dot{M}\rangle$ by evolving through a large number of H shell flashes (See discussions in \cite{Townsley_Bildsten_2004}). However, once the WD core equilibrate, the observable quantities remain largely insensitive to $M_{\rm WD}$ and $\langle\dot{M}\rangle$. Figure \ref{fig:TemperatureprofiledifferentTc} presents the temperature profiles of a 0.78~$M_\odot$~WD as a function of fractional depth ($\log(1-m/M)$). The profiles are shown for three core temperatures, $T_{\rm c} = 5,7$ and $9\times10^6$ K represented with red, blue, and green lines, respectively. The top panel shows the temperature profiles at the beginning of the long-term accretion phase (solid lines) and when the surface hydrogen luminosity reaches $L_{\rm H}=10^{2}~L_\odot$ (dashed lines). The bottom panel display the profiles just before shell's ignition or pre-outburst (solid lines, onset of strong convection) and slightly advanced phase of the WD evolution (well past the shell's ignition point), when $L_{\rm H}=10^{5}~L_\odot$ (dotted lines). From the innermost region of temperature profile shown in Figure \ref{fig:TemperatureprofiledifferentTc}, we see that the WD center experience significant cooling for the $T_{\rm c}=9\times 10^6$~K model during the accumulation of the surface layer and remains in overall net cooling phase even as ignition approaches. In this case, the WD exhibits an outward temperature gradient between $\log(1-m/M) = -2$ and -1. However, for the lower temperature model, $T_{\rm c}=5\times 10^6$~K, the WD initially cools but subsequently the temperature profile changes enough to stop the cooling of or possibly slightly reheat the center as $M_{\rm acc}$ increases. For this model, the temperature gradient becomes inward over the same fractional mass depth of -2 to -1. This feature suggests that the WD equilibrium core temperature should lie between 5 and $7\times10^7$~K, where core cooling is balanced by the heating for a given $\langle\dot{M}\rangle$ across the CN cycle. However, running a full CN evolutionary model with both overshooting and element diffusion enabled in \texttt{MESA} does not appear to be currently feasible, as they produce nonphysical profiles due to the various mixing processes in place. This numerical artifact is particularly evident in the bottom panel of Figure \ref{fig:TemperatureprofiledifferentTc} for the $T_{\rm c} = 5\times 10^6$~K. Due to these limitations, we are currently unable to perform a full CN cycle that includes both overshooting and element diffusion. 

We construct five different non-magnetic CV WD models with $T_\text{c} = 5, 5.5, 6, 6.5, \text{and}~7 \times 10^{6}$~K for a 0.78~$M_\odot$~WD. In order to determine the approximate ignition mass, the different $T_c$ models are subjected to long-term accretion with $\langle\dot{M}\rangle=6\times 10^{-11}M_\odot$~yr$^{-1}$, resulting in ignition masses of $M_\mathrm{ign}=3.57, 3.05, 2.84, 2.05,$ and  $1.69\times10^{-4}M_\odot$ for $T_{\rm c}=5,5.5,6, 6.5,$ and  $7\times 10^{6}$~K, respectively for the 0.78~$M_\odot$ model. To ensure consistency, we select models with the same accreted mass $M_{\rm acc} = 1.5 \times 10^{-4}M_\odot$ for all $T_{\rm c}$ values, as this mass remains below the $M_{\rm ign}$ in every case. At this $M_{\rm acc}$, we determine an appropriate $\langle\dot{M}\rangle$ that yields the star surface temperature of $T_{\rm eff} = 14000$~K, for all $T_{\rm c}$ values. We find  $\langle\dot{M}\rangle = 12.37, 11.42, 11.10, 9.42,$ and $7.94 \times 10^{-11} M_\odot ~ {\rm yr}^{-1}$, respectively for $T_{\rm c} = 5, 5.5, 6, 6.5,$ and $7\times 10^{6}$~K. Some of these $T_{\rm c}$ are low enough that the plasma Coulomb parameter $\Gamma$ exceeds the threshold for forming a solid \citep[$\Gamma \gtrsim 175$,][]{Isern_1997, Potekhin_chabrier_2000, Potekhin_chabrier_2010}. Consequently, a fraction of the WD core solidifies, with the extent of crystallization depending on both $T_{\rm c}$ and the WD mass, $M_{\rm WD}$. For a 0.78~$M_\odot$~WD, the solid core fraction contributes approximately 26 (34), 12 (25), and 3 (15)\% cores that are solid in mass (radius) coordinates, for $T_{\rm c} = 5, 5.5,$ and $6 \times 10^{6}$~K. However, at higher core temperatures $T_{\rm c} = 6.5$ and $5.5\times 10^{6}$~K, the WD core remains uncrystallized.

Figure \ref{fig:Abundanceandbrunt0.78M5e67e6K} illustrates the WD thermal structure following the long-term accretion phase with $M_{\rm acc} = 1.5\times 10^{-4}M_\odot$ for $T_{\rm c}=5 ~(solid ~lines)$ and $7\times 10^{6} ~(dashed~lines)$~K for a 0.78~$M_\odot$ at $T_{\rm eff}=14000$~K. The sloped abundance profiles in the H-rich layer reflect the combined effects of thermohaline mixing and element diffusion operating throughout the accretion phase. The top two panels display the mass fractions of the key elemental species, with the middle panel shown on a logarithmic scale to display the significant contributions from the other less abundant but important species. The vertical dashed line (black) indicates the $\log P$ at which the core is completely solid for $T_{\rm c}= 5\times 10^6$~K. The bottom panel presents the buoyancy and lamb frequency profiles. The peak in buoyancy frequency at $\log P  \approx 19$ $\mathrm{erg ~cm^{-3}}$ reflects the transition between the hydrogen-rich layer and helium. Another peak located between $\log P$ of $\sim$~23 and 24 is contributed due to the CO gradient at the edge of the remnant of the convective core formed during core He burning. Notably, the overall neon abundances remain the same irrespective of WD core temperatures. The slight discontinuity at $\log P$ of $\sim 15.8$ corresponds to the outermost layer in which thermohaline mixing is effective due to the finite timestep in the simulation. While the presence or absence of a solid core does not lead us to any obvious differences in the internal structural features, asteroseismology may be able to provide insights into the extent of core crystallization through the forward modeling method, as we continue to study here.
\begin{figure}[ht]
    \centering
    \includegraphics[width=1.\columnwidth]{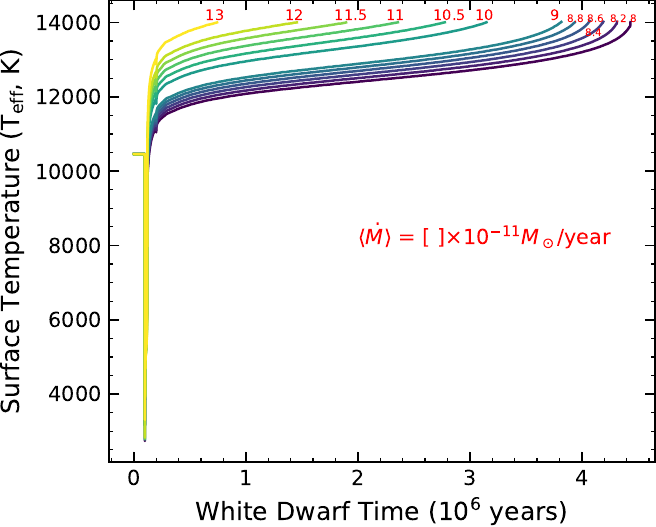}
    \caption{Evolution of the WD surface temperature during the long-term accretion phase is shown for a 0.78~$M_{\odot}$ model. The figure presents 12 different models with $\langle\dot{M}\rangle $~(right to left): $8,8.2, 8.4, 8.6, 8.8, 9, 10, 10.5, 11, 11.5, 12$, and $13 \times 10^{-11}~ M_\odot~{\rm yr}^{-1}$. Each model is characterized by a $T_{\rm c} = 5\times 10^{6}$~K and a final effective temperature $T_{\rm eff} = 14000$~K. The minimum and maximum $\langle\dot{M}\rangle$ values end at $M_{\rm acc}$ of 3.467 and $0.837 \times 10^{-4}~M_\odot$, respectively. The highest $\langle\dot{M}\rangle$ results in the WD reaching $T_{\rm eff} = 14000$~K in the shortest timescale, within in approximately one million years.}
    \label{fig:accrateTeff14000}
\end{figure}

\subsubsection{Varying Accreted Mass}\label{sssec:varyaccretedmass}
In the second modeling scenario, we investigate variations in $M_{\rm acc}$ while keeping $T_{\rm c}$ and $T_{\rm eff}$ fixed. Specifically, we evolve models across 12 different $\langle\dot{M}\rangle$ values with $8,8.2,8.4, 8.6, 8.8, 9, 10, 10.5, 11, 11.5, 12$, and $13 \times 10^{-11}M_\odot~{\rm yr}^{-1}$. The minimum and maximum $\langle\dot{M}\rangle$ values result in  $M_{\rm acc}$ of 3.467 and 0.837~$\times 10^{-4}M_\odot$, respectively, with other $M_{\rm acc}$ values detailed in Table \ref{table:table_all_parameters}. This is consistent with implications drawn by \cite{Townsley_Bildsten_2004}, with a higher $M_{\rm acc}$ implying a lower $\langle\dot{M}\rangle$ for the same $T_{\rm eff}$. These models correspond to a 0.78~$M_\odot$ WD with $T_{\rm c} = 5\times 10^{6}$~K and $T_{\rm eff} = 14000$~K, where approximately 26\% of WD mass (33 - 34\% by radius) is solid at this $T_{\rm c}$. The time evolution is shown in Figure \ref{fig:accrateTeff14000} with all WD models reaching the same final effective temperature of 14000~K. Although the accreted envelope of the WD may slightly modify the star's deep thermal structure, specifically $T_{\rm c}$, it remains largely insensitive to $\langle\dot{M}\rangle$.

\subsubsection{Varying WD Mass}\label{sssec:varywdmass}
In the third modeling scenario, we consider models with five different masses $M_{\rm WD}$: 0.60, 0.70, 0.78, 0.86, and 0.93~$M_\odot$. The choice of WD mass falls within the range of observed CV masses \citep{Pala_2022}. By considering WD mass within 0.6-0.93~$M_\odot$, we effectively capture the diversity of CV systems and their evolutionary stage. Each model has fixed $T_{\rm c} = 6\times 10^{6}$~K. At this $T_{\rm c}$, the WDs with masses of 0.78, 0.86, and 0.93~$M_\odot$ have solid cores approximately 3 (15), 13 (25), and 33 \% (37\%) in mass (radius), respectively. In contrast, the WDs with masses of 0.6 and 0.7~$M_\odot$ have their cores entirely in the liquid phase. For each WD mass, we then select an accreted envelope mass equal to two-thirds of $M_{\rm ign}$, corresponding to 2.55, 2.08, 1.89, 1.38, and 1.01$\times 10^{-4}M_\odot$ for 0.6, 0.7, 0.78, 0.86, and 0.93~$M_\odot$, respectively. These choices yield $\langle\dot{M}\rangle$ = 12.75, 10.43, 8.34, 7, and 5.23 $\times~10^{-11}M_\odot$~yr$^{-1}$, ensuring that all models reach a final effective temperature of $T_{\rm eff} = 14000$~K. The $M_{\rm ign}$ decreases as the $M_{\rm WD}$ increases at a fixed $\langle\dot{M}\rangle = 6\times 10^{-11}M_\odot$~yr$^{-1}$. The parameter space is summarized in Table \ref{table:table_all_parameters}.

\begin{table*}[ht]
    \centering
    \caption{WD models and parameter space}
    \begin{tabular}{c|c|c|c|c|c|cc}
    \hline \hline
    WD Mass & $T_{\rm c}$ &Extent of Core & Extent of Core& $\langle\dot{M}\rangle$ & $M_{\rm acc}$\footnote{All models with the final surface temperature $T_{\rm eff} = 14000$ K} & $M_{\rm ign}$ at\\
    ($M_\odot$)& (10$^6$~K) & Crystallization& Crystallization & (10$^{-11}$~$M_\odot$/yr) & (10$^{-4}$~$M_\odot$)& $\langle\dot{M}\rangle = 10^{-10.22}$~$M_\odot$/yr \\
    & & (\% Radius)& (\% Mass)& & & (10$^{-4}$~$M_\odot$)\\
     \hline \hline
     \multirow{5}{*}{0.78}& 5.0 & 34 &26 & 12.37 & 1.50 & 3.57 \\ \cline{2-7}
    &  5.5 & 25 & 12 & 11.42 & 1.50& 3.05 \\ \cline{2-7}
    &  6.0 & 15 & 3 & 11.10 & 1.50& 2.84  \\ \cline{2-7}
    & 6.5 & No Solid Core& No Solid Core & 9.42 & 1.50& 2.05  \\ \cline{2-7}
    & 7.0 & No Solid Core& No Solid Core & 7.94 & 1.50& 1.69 \\ \cline{2-7}
    \hline \hline
    & 5.0 & 33& 26 & 8.0 & 3.467& - \\ \cline{2-7}
    & 5.0 & 33& 26 & 8.2 & 3.455& - \\ \cline{2-7} 
    & 5.0 & 33& 26 & 8.4 & 3.437& - \\ \cline{2-7} 
    & 5.0 & 33& 26 & 8.6 & 3.414& - \\ \cline{2-7} 
    & 5.0 & 33& 26 & 8.8 & 3.375& - \\ \cline{2-7}
    & 5.0 & 33& 26 & 9.0 & 3.344& - \\ \cline{2-7}
0.78& 5.0 & 33& 26  & 10 & 3.048& - \\ \cline{2-7} 
    & 5.0 & 33& 26  & 10.5 & 2.81& - \\ \cline{2-7} 
    & 5.0 & 33& 26 & 11 & 2.483& - \\ \cline{2-7}
    & 5.0 & 33& 26  & 11.5 & 2.064& - \\ \cline{2-7} 
    & 5.0 & 34& 26 & 12 & 1.627& - \\ \cline{2-7}
    & 5.0 & 34& 26 & 13 & 0.837& - \\ \cline{2-7}
    \hline \hline
    0.60& 6.0 & No Solid Core& No Solid Core& 12.75 & 2.55& 3.825 \\ \hline 
    0.70& 6.0 & No Solid Core& No Solid Core & 10.43 & 2.08& 3.12 \\ \hline 
    0.78& 6.0 & 15& 3 & 8.34 & 1.89& 2.835 \\ \hline 
    0.86& 6.0 & 25& 13 & 7.0 & 1.38& 2.07 \\ \hline 
    0.93& 6.0 & 37& 33 & 5.23 & 1.01& 1.515\\ \hline 
   \hline
    \end{tabular}
    \label{table:table_all_parameters}
\end{table*}

\subsection{Choice of Modes} \label{ssec:choiceofmodes}
The associated pulsations in accreting WDs are generally attributed to be $g$ modes similar to those commonly applied to the isolated WDs. These $g$ modes are excited thermally by the convection zone related to hydrogen and helium ionization zones, as in DAV and DBV stars \citep{Brickhill_1991, Arras_Townsley_Bildsten_2006, Saio_2013}. However, some recent studies in the literature suggest that these pulsations could be Rossby modes \citep{Saio_2019}. In our companion paper, we conducted a simultaneous analysis of both $g$ and $r$ modes, demonstrating that within the currently established framework of the convective driving mechanism, the pulsations in accreting WDs should be expected to have a $g$-mode origin \citep{Kumar_Townsley_2026}. We explicitly computed the visibility amplitudes of both $g$ and $r$ modes and highlighted that while a few retrograde $r$ modes exhibit larger visibility, the low-order $g$ modes possess higher frequencies in the star's frame, thus making them more likely to be driven in the star's frame and observable. 

Following the findings of \cite{Kumar_Townsley_2026}, with the investigation of the surface amplitudes of $g$ modes in accreting WDs, we focus exclusively on $g$ modes up to the second-lowest order ($\ell=2$) and disregard the $r$ mode significance in such systems in this present analysis. We use \texttt{GYRE} to compute adiabatic $g$-mode frequencies just as described in \cite{Kumar_Townsley_2026}.

\section{Results}\label{sec:results}
\begin{figure*}[ht]
  \includegraphics[scale =0.9]{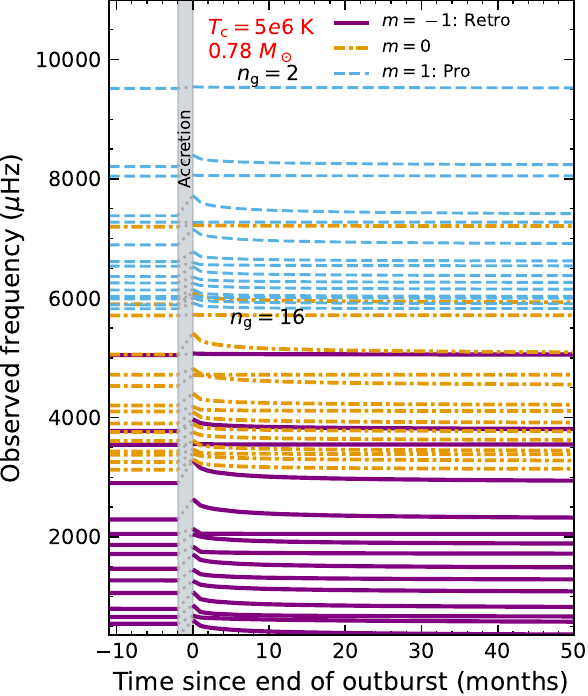}
  \includegraphics[scale= 0.9]{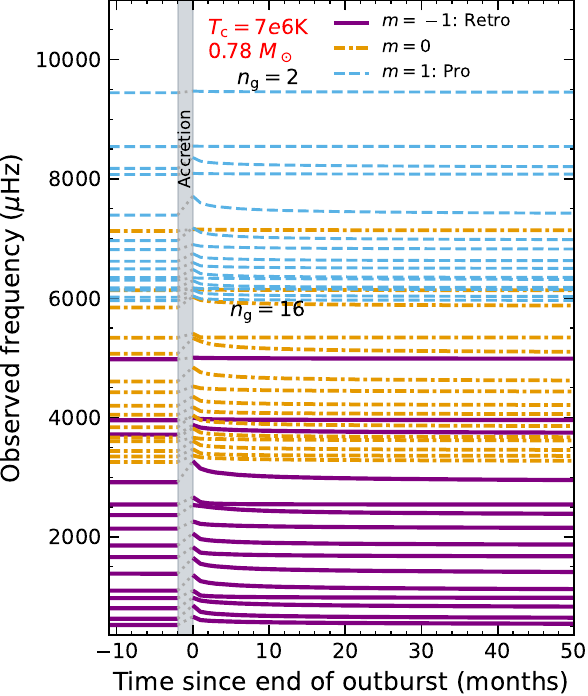}
   \caption{Gravity mode frequencies in the inertial frame against time (in months) since the accretion outburst for 0.78~$M_\odot$ mass model. The left panel is for $T_{\rm c} = 5\times 10^6$~K (crystallized core), and the right panel is for $T_{\rm c} = 7\times 10^6$~K, both for a rotation period of 209 s. Top to bottom indicates $n_{\rm g}=2–16$. Three different colors and line styles indicate the three values of the azimuthal eigenfunction index $m$ that make up the dipole triplets.}
   \label{fig:0.78Mtimevariation} 
\end{figure*}
We now present our seismological calculations for the accreting WD models described in section \ref{sec:method}. We consider the $g$ modes in the rapid rotation limit using the observed spin period of GW Lib, 209 s, as a representative value for CV systems \citep{Szkody_2012}. This period is much shorter than the hours or days observed in isolated WDs. In this regime, rotation starts to strongly and nonlinearly modify $g$-mode frequencies once the spin parameter $2\Omega/\omega \gtrsim 1$. Given that $g$-mode periods in GW Lib are comparable to its spin period, the system lies within the regime where rotational effects are significant and hence the Traditional Approximation of Rotation (TAR) is needed. Although, \texttt{GYRE} allows the treatment of rotation both with and without the Cowling approximation \citep{Cowling_1941}, we conduct analysis within the TAR and adopt the Cowling approximation, consistent with common practice in the literature \citep{Bildsten_Ushomirsky_Cutler_1996, Lee_Saio_1987, Townsend_2020}. By disabling it in GYRE, we find that $g$-mode frequencies will change by about 0.002\% if the Cowling approximation is not used. We present $g$-mode frequency variation with $T_{\rm c}$ in section \ref{ssec:gmodetcore}, quantifying the mode frequencies with and without a partial solid core. Furthermore, we propose a potential new method to identify the mode orders based on the time variation of frequencies during the months of cooling after the strong accretion event. Section \ref{ssec:gmodethicknesslayer} introduces the variation of the $g$-mode frequencies with the accreted layer of the star, $M_{\rm acc}$. Finally, we show the changes for the different mass models.

\subsection{Gravity Modes vs. Core Temperatures} \label{ssec:gmodetcore}
Here we begin by showing the $g$-mode eigenfrequencies following the strong accretion event. These strong accretion outbursts occur on relatively short timescales compared to the overall age of the star. However, the intermittent accretion events with finite quiescence periods can significantly disrupt the outer layers, leading to modifications in the $g$-mode frequencies. We investigate the radial mode orders for the dipole mode ($\ell = 1$) after the strong accretion event, tracking the evolution of mode frequencies over a period of just over four years. These calculations build upon the work of \cite{Kumar_Townsley_2023}, but with a significantly improved WD model as outlined in the section \ref{ssec:longtermaccretion}.   
Figure \ref{fig:0.78Mtimevariation} illustrates the time variations of the $g$-mode frequencies for radial orders $n_{\rm g} = 2 -16 ~(top ~ to ~ bottom)$ in the observer's frame of a dipole mode, following the accretion outbursts for a 0.78~$M_\odot$~WD model. The left panel corresponds to $T_{\rm c} = 5 \times 10^6$~K, where the WD core has a solid core, while the right panel shows results for $T_{\rm c} = 7 \times 10^6$~K. The dashed blue lines are the prograde modes ($m = 1$), the purple solid lines are the retrograde modes ($m=-1$), and the orange dotted-dashed lines are the zonal modes. These are evaluated in the inertial frame of reference and so correspond to the frequencies observed through the brightness variations by a distant observer. The gray shaded region indicates the two months of strong accretion during the DNe outburst. The minus values of the time axis represent the time before the strong accretion event. 

Both the panels exhibit relatively similar trends, as each radial mode order frequency gradually relaxes to its preoutburst value within approximately 36 months. The sudden temperature rise in the outer layers due to the accretion event causes the mode frequencies to rise by a few percent relative to their quiescent value. However, the lowest order $g$ modes (for example, $n_{\rm g} = 2$) are somewhat less affected by the temperature change in the accreted envelope as they reside much deeper into the star, making their mode frequencies less sensitive to heating in the outer layers. While the overall qualitative behavior of mode relaxation remains consistent and similar with our earlier work \citep{Kumar_Townsley_2023}, the WD models presented in this work incorporate significant improvements. The presence of element diffusion during the long-term accretion phase leads to smoother buoyancy profiles between the accreted material and the WD core. Additionally, our treatment of CV evolution is relatively realistic with adjusted core temperatures to reflect the observational constraints, hence giving a more correct mode frequency within the star.
\begin{figure}[ht]
  \includegraphics[width=1.\linewidth]{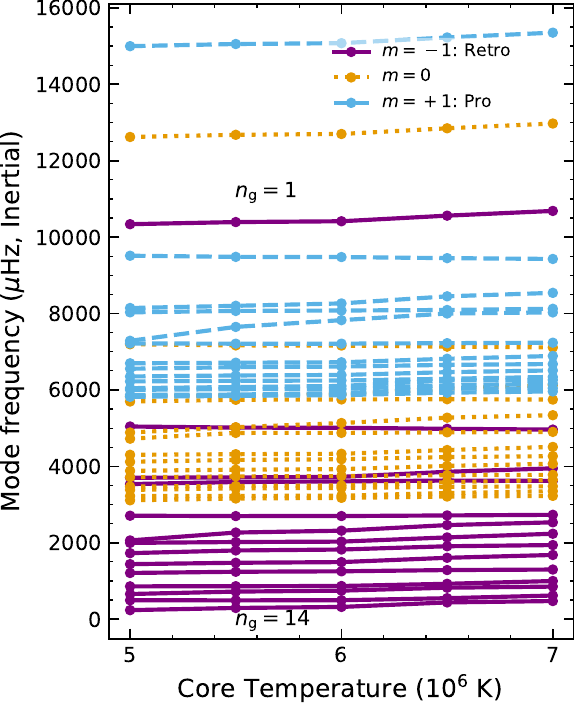}
   \caption{First 14 $g$-mode frequencies in the inertial frame are shown as a function of $T_{\rm c}$ calculated for the dipole mode for a 0.78~$M_\odot$ WD model following the long-term accretion phase. The models have an accreted mass of $M_{\rm acc} = 1.5\times 10^{-4}~M_\odot$. Solid purple lines are the retrograde modes, dashed blue lines are the prograde modes, and dotted orange lines are the zonal modes ($m=0$).}  
   \label{fig:0.78coretempvariation} 
\end{figure}

Figure \ref{fig:0.78coretempvariation} investigates the $g$-mode frequencies observed by an observer as a function of the WD core temperature ($T_{\rm c}$) for a 0.78~$M_\odot$ model. Three of the models at $T_{\rm c}$ = 5, 5.5, and 6$\times 10^6$~K have a solid core, with each having a different size. All the models share the same final $T_{\rm eff}$ and an accreted mass of $M_{\rm acc} = 1.5\times 10^{-4}~M_\odot$. As shown in the figure, the low-order (highest frequency, shortest period) prograde $g$ modes (dashed blue lines) exhibit notable changes in presence of a solid core, highlighting the influence of core crystallization on observed mode frequencies. Although, in this work, we do not study the $g$-mode structure with the moving crystallization front; rather we just focus on mode behavior with the effect of solid cores. The presence of a partial solid core relocates the inner boundary of the mode propagation, moving the solid-liquid interface of the star, thereby reducing the resonant cavity of the core. We apply the hard-sphere boundary condition at the inner solid/liquid interface, where the radial displacement is set to zero \citep{Montgomery_1999}. The extent of crystallization at lower $T_{\rm c}$ pushes these modes farther out from the center. 
\begin{figure}[h]
    \centering
    \includegraphics[width=1\linewidth]{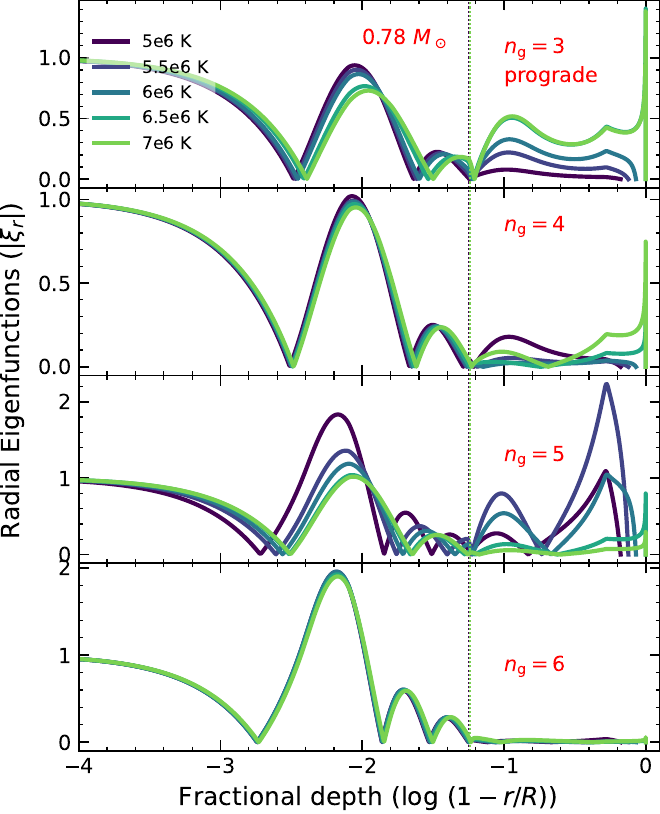}
    \caption{Absolute radial eigenfunctions ($|\xi_r|$) for prograde modes with radial orders $n_{\rm g} = 3, 4, 5,$ and 6, as a function of the fractional depth into the star ($\log (1-r/R)$). The eigenfunctions are computed using Gyre for different core temperatures (as labeled) of a 0.78~$M_\odot$~WD model with $M_{\rm acc} = 1.5\times 10^{-4}~M_\odot$. The two vertical dashed lines indicate the boundary between the core and the envelope for $T_{\rm c} = 5\times 10^6$~K (left) and $T_{\rm c} = 7\times 10^6$~K (right). The amplitudes are normalized so that all the modes have the same surface amplitudes. The core of the star is the extreme right of the plot.}
    \label{fig:Radialeigenfunctions}
\end{figure}

Remarkably, prograde mode orders  $n_{\rm g}=$ 3, 4, 5, and 6 exhibit pronounced variations with core temperatures, potentially showing stronger influence due to the inner solid-liquid interface. Figure \ref{fig:Radialeigenfunctions} illustrates the absolute radial eigenfunction, $|\xi_r|$, for these modes ($n_{\rm g}=$ 3, 4, 5, and 6) as a function of the logarithm of the fractional depth into the star, $1-r/R$, at different $T_{\rm c}$ for a 0.78~$M_\odot$~WD model. The amplitudes are normalized such that all modes have the same surface amplitudes, in this case 1. The node locations shift modestly with core temperature, with the displacement being more evident for $n_{\rm g}=5$. The core (right side of the figure) and envelope cavity are indicated with the vertical dashed lines for $T_{\rm c} = 5\times 10^6$~K (left) and $T_{\rm c} = 7\times 10^6$~K (right). The mode behavior in the core and envelope is particularly important. These radial perturbations of the modes are highly influenced by the way Brunt-V\"ais\"al\"a and Lamb frequencies allow them to propagate within the star, as even the slightest variation in the Brunt-V\"ais\"al\"a could potentially alter the mode propagation within the star. The buoyancy frequency is set by the gradients produced by the temperature, density, and compositions within the star. As the core temperature increases, these gradients shift the node locations significantly. While the overall node locations remain largely unchanged in the shell (H-rich) for even order modes ($n_{\rm g}= 4, 6$), they are distinctly positioned in the core H-poor cavity. As a general trend, the overall perturbation amplitudes of the mode (both even and odd) in the shell decreases as $T_{\rm c}$ increases, with the odd-order modes exhibiting larger variations. However, this feature is switched in the core with some mode order ($n_{\rm g}=3$) showing stronger perturbation amplitude at higher $T_{\rm c}$ and other ($n_{\rm g}=5$) carrying the same behavior as in the shell. At lower $T_{\rm c}$, the inner node of $n_{\rm g}=4$ is located relatively closer to core-shell boundary and the amplitudes are smaller than the $n_{\rm g}=3$ mode compared to higher $T_{\rm c}$, elucidating the mode spacing in Figure \ref{fig:0.78coretempvariation}. Comparatively, the $n_{\rm g}=5$ mode shows a larger core variation with $T_{\rm c}$ than $n_{\rm g}=6$ mode, with the lower $T_{\rm c}$ producing stronger core perturbations implying mode confinement in the core. This behavior is also noted in DA WDs and sdB stars \citep{Brassard_1992, Charpinet_2000}.
\begin{figure}[h]
    \centering
    \includegraphics[width=1.\linewidth]{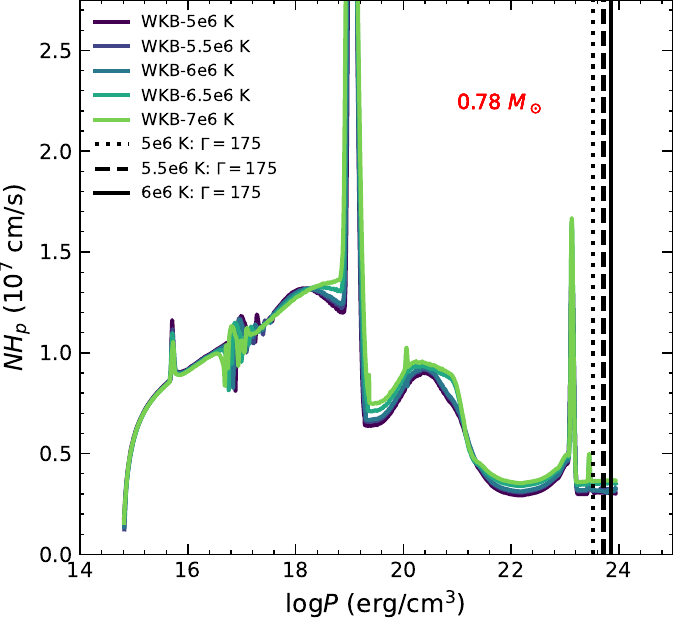}
    \caption{WKB integrand for the core temperatures $T_{\rm c} = 5,5.5,6,6.5,$ and $7 \times 10^6$~K for a 0.78~$M_\odot$~WD model. The outer boundary of the solid core is indicated by the black lines, taken at $\Gamma = 175$. The core and envelope boundary is located at $\log (P/\rm erg~cm^{-3}) \sim 19$.}
    \label{fig:wkbcoretemp}
\end{figure}
\begin{figure}[h]
  \includegraphics[width=1.\linewidth]{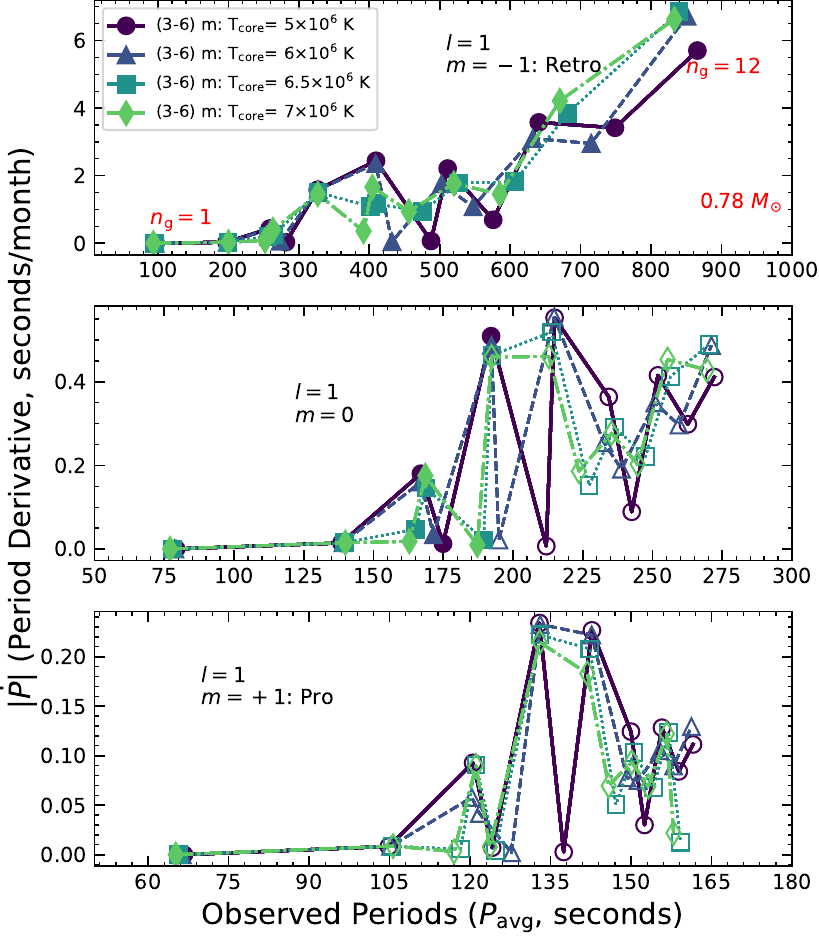}
   \caption{The rate of period change as a function of average observed periods of mode orders $n_{\rm g} = 1-12$, following the strong intermittent accretion aka Dwarf Nova cycle for the several core temperatures (as labeled in the top panel) of a 0.78~$M_\odot$ model with an accreted layer mass of $1.5\times 10^{-4}~M_\odot$. The rates are evaluated at 4.5 months after the outburst. The dipole ($\ell=1$) modes, retrograde, zonal, and prograde modes are displayed in the top, middle, and bottom panels, respectively. Closed symbols indicate for the modes that are driven in the star's frame and therefore likely to be observed. The minimum driving mode period is taken as 192 seconds, in agreement with our earlier work \citep{Kumar_Townsley_2026}.}
   \label{fig:PdotPforcoretemp0.78M} 
\end{figure}

To assess these effects concretely, we often focus on the WKB integrand, which provides a direct measure of the mode frequency variations rather than just the buoyancy frequency alone. The integrand reflects any localized structural changes within the star and is proportional to the number of nodes present per decade in pressure inside the star. Within the WKB approximation, a plane wave form is assumed with a radius-dependent wavenumber $k_r$. Consequently, the phase accumulation, $\phi$, of the wave in a given region is $\phi$ = $\int$ dr $k_{r}=[\ell(\ell+1)]^{1/2}\omega^{-1}\int N\;dr/r$ where $N$ is the Brunt-V\"ais\"al\"a frequency, $\ell$ is the angular order of the mode, and $\omega$ is the frequency of the mode. For a mode of fixed radial order, solving for $\omega$ yields, approximately, $\omega \propto (1/R) \int N H_{P}\;d\ln P$, where $H_P$ is the pressure scale height. We identify $N H_P$ as the WKB integrand and note that a plot of $N H_P$ vs. $\log P$ can be integrated by eye to convey how various features inside the star contribute to changes in $\omega$.

To quantify the changes in the mode frequencies, figure \ref{fig:wkbcoretemp} represents the WKB integrand for the different core temperatures ($T_{\rm c} = 5,5.5,6,6.5,$ and $7 \times 10^6$~K) of a 0.78~$M_\odot$~WD model following the long-term accretion with $M_{\rm acc} = 1.5\times 10^{-4} M_\odot$. The boundary of the crystallized core is indicated with the black solid, dashed, and dotted lines for $T_{\rm c} = 5,5.5,$ and $6\times 10^6$~K, respectively, determined by the condition $\Gamma=175$. The transition between the core and the envelope occurs at $\log (P/\rm erg~cm^{-3}) \sim 19$. The relative strength of the WKB integrand (computed as envelope to core ratio) decreases from 1.5 at $T_{\rm c} = 5\times 10^{6}$~K to 1.25 at $T_{\rm c} = 7\times 10^{6}$~K. As a result of this change in contrast with $T_{\rm c}$, modes that have their dominant amplitude in the core with their frequencies shift compared to those which have their dominant amplitude in the envelope. This contrast matches well with the comparison of eigenfunctions in Figure \ref{fig:Radialeigenfunctions}, in which the envelope mode $n_{\rm g}=6$ is fairly unchanged, whereas the modes with more core amplitude, $n_{\rm g} = 3$ and 5, have their eigenfunctions and mode frequencies change significantly. This may make it possible to constrain $T_{\rm c}$ and, along with it, crystallized fraction, since both of these contribute to the change in the WKB integrand in the inner part of the star.

Identifying the true mode orders in accreting WDs is challenging due to the limited number of observed modes. A common approach is to infer the radial order mode by identifying the consecutive radial order modes and matching them with the theoretical models within the reasonable parameter space. Here, we propose a potential new method for mode identification based on the time-variation of period or frequency during the months of cooling after the accretion event. Figure \ref{fig:PdotPforcoretemp0.78M} presents the period derivative ($|\dot{P}|$) as a function of average observed periods for mode orders $n_{\rm g} = 1-12$ for the dipole ($\ell=1$) modes, following the strong intermittent accretion event for $T_{\rm c} = 5, 6, 6.5$, and $7\times10^6$~K of a 0.78~$M_\odot$ model with $M_{\rm acc} = 1.5\times 10^{-4}~M_\odot$. The period derivative is evaluated at 4.5 months post DN outbursts. A fiducial shortest driven mode period of approximately 192 seconds is adopted in star's frame, rather than deriving it from a full non-adiabatic mode calculation; this value is representative of what may be appropriate for GW Lib. Filled symbols indicate modes that are likely to be observed. Notably, retrograde modes exhibit the largest $|\dot{P}|$, approximately an order of magnitude higher than prograde and zonal modes. By observing a specific mode and computing its period derivative, this could reveal the true mode order without examining the neighboring modes. This technique may offer a valuable tool for mode identification.

\begin{figure}[ht]
  \includegraphics[width=1.\linewidth]{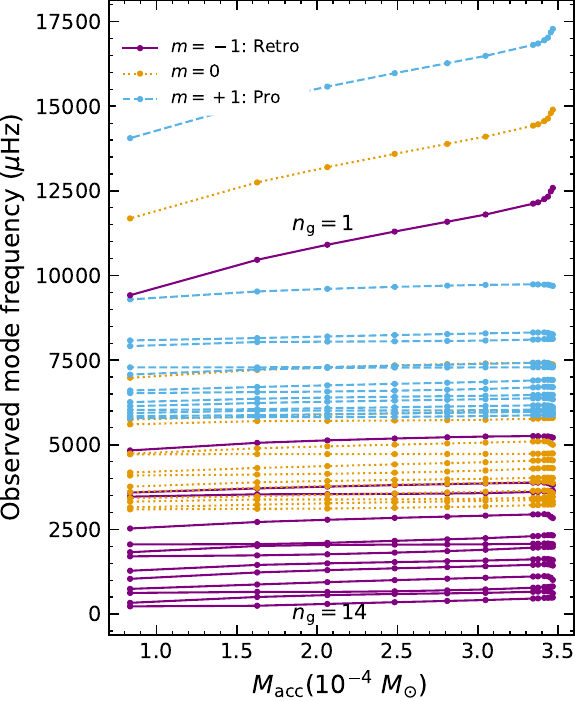}
  
   \caption{First 14 $g$-mode frequencies (top to bottom) against the accreted layer mass, $M_{\rm acc}$, for the dipole mode of a 0.78~$M_\odot$~WD model. Prograde, zonal, and retrograde modes are shown with dashed blue, dotted orange, and solid purple lines, respectively.}
   \label{fig:modesvsmacc0.78M} 
\end{figure}
\subsection{Gravity Modes vs. Accreted Layer Thickness} \label{ssec:gmodethicknesslayer}
Now, we explore the dependence of mode frequency on the accreted layer mass, $M_{\rm acc}$, which varies during the long-term accretion before the onset of the classical novae. The direct measurement of the accreted layer mass is not feasible by means other than seismology; however, the amount of mass ejected during the classical novae events has been constrained for some related systems \citep{Townsley_2005}.

\begin{figure}[h]
    \centering
    \includegraphics[width=1.\linewidth]{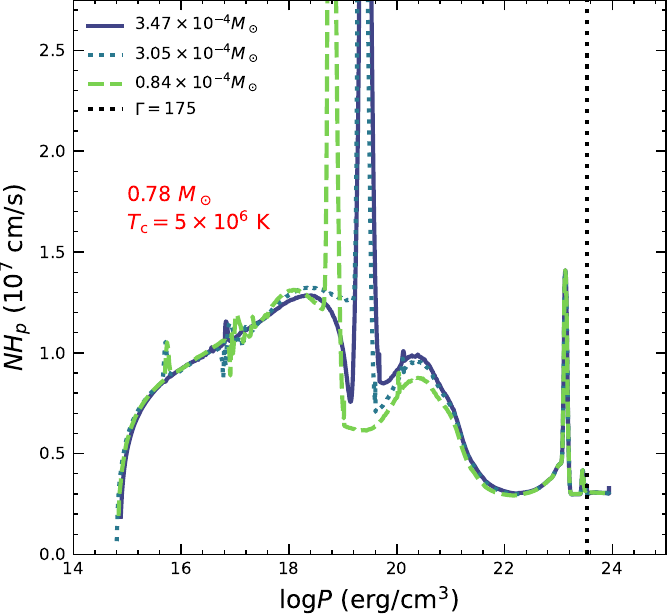}
    \caption{WKB integrand at three different accreted thickness layer, $M_{\rm acc} = 0.84,3.05$ and $3.47\times 10^{-4}~M_\odot$, for a 0.78~$M_\odot$~WD model with $T_{\rm c} = 5\times 10^6$~K.}
    \label{fig:wkbthickness}
\end{figure}

Figure \ref{fig:modesvsmacc0.78M} investigates the observed mode frequency variation of the first 14 radial orders (top to bottom) with the accreted layer mass, $M_{\rm acc}$, for a 0.78~$M_\odot$~WD model with $T_{\rm c} = 5\times 10^6$~K. In general, mode frequencies increase with $M_{\rm acc}$. The solid purple, dashed blue, and dotted orange lines represent retrograde, prograde, and zonal modes for the $\ell =1$ triplets, respectively. The overall mode frequency increases with the newly accreted layer. The actual reason for this change is shown in Figure \ref{fig:wkbthickness} - the Brunt, and therefore the WKB integrand, is higher in the envelope, so that as $M_{\rm acc}$ increases, the area under the integral curve shown in Figure \ref{fig:wkbthickness} increases. This leads to higher mode frequencies since the mode frequencies are proportional to this integral, as discussed in the previous subsection. This enhancement arises both from hydrogen to helium transition in the Brunt-V\"ais\"al\"a frequency and from the structural changes to the outer cavity associated with the newly accreted layer. While the $n_{\rm g}=1$ mode exhibits the largest change and continuous increase with $M_{\rm acc}$, the other radial order modes show a slight dip when $M_{\rm acc}$ approaches the $M_{\rm ign}$. The evolving accreted envelope effectively breaks the star into three distinct cavities: core, shell, and transition layer. The transition layer between the core and the shell plays a crucial role in determining node locations for different mode orders. While the mode orders $n_{\rm g} >1$ warrant the existence of at least one node either in the shell or core, the single node in the $n_{\rm g}=1$ mode constrained near the transition layer, making it highly sensitive to variations in the newly accreted layer.  

\begin{figure}[h]
  \includegraphics[width=1.\linewidth]{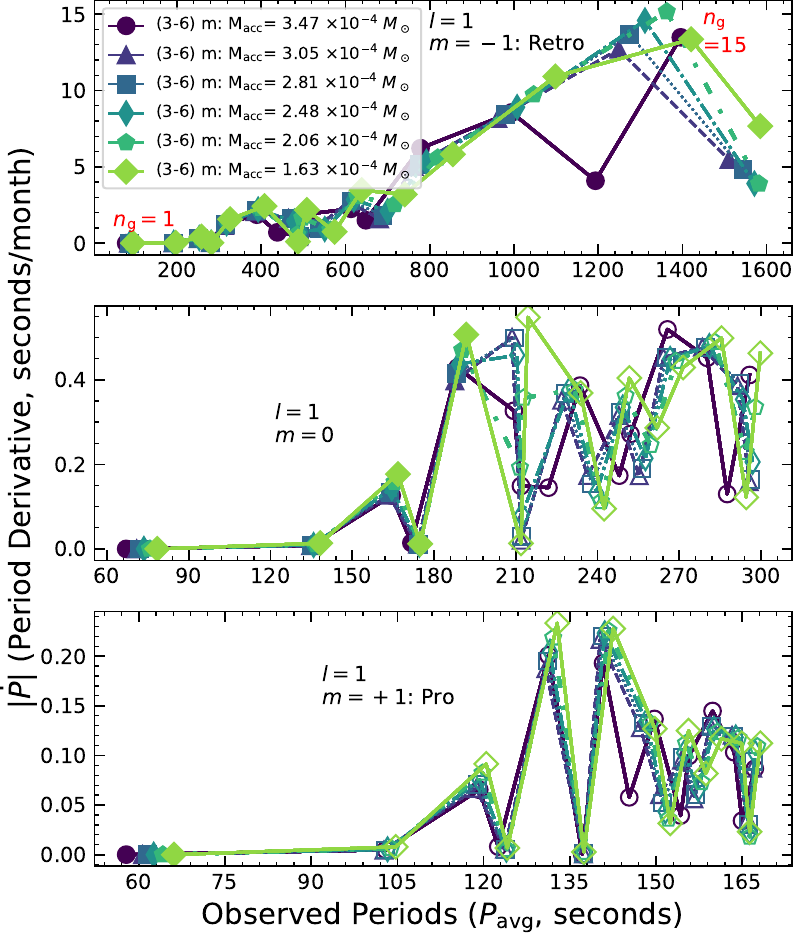}
   \caption{Variation of the observed gravity mode frequencies against the WD accreted mass layer, $M_{\rm acc}$ at the effective temperature of 14000 K of a 0.78~$M_\odot$ model.}
   \label{fig:PdotPforMaccc0.78M} 
\end{figure}
A striking feature that is quite evident in Figure \ref{fig:modesvsmacc0.78M} is mode pairing (alternating small-large frequency spacing). This is particularly noticeable in the both prograde and retrograde mode orders, such as prograde $n_{\rm g} = 3-4$, $5-6$ and retrograde $n_{\rm g}= 6-7$. As the radial order increases, a node is allocated either in the core or the envelope, depending on the mode. This leads to alternation between the smaller and larger frequency spacings between the consecutive modes. In a uniform stellar structure, $g$ modes are expected to have inverse frequencies which are uniformly spaced, with higher-order modes appearing closer together. However, in WDs, the contrast between the light-element envelope and the heavier-element core disrupts this regularity making the spacing irregular \citep{Townsley_Arras_Bildsten_2004, Fontaine_2008, Althaus_2022}. In our accreting WD, there is a single localized division between the accreted light-element (H+He) layer and the inner CO material, since we have hypothesized that the entire envelope will be ejected and then re-accreted. The thickness and structure of the subsurface layer modify the density gradient, therefore shifting the Brunt-V\"ais\"al\"a frequency profile. These changes are significantly effective in the transition region between the core and envelope, where the mode confinement and trapping conditions are set. The modification to Buoyancy frequency appears directly in the WKB integrand, which determines the strength of the cavities. 

Figure \ref{fig:wkbthickness} examines the WKB integrand for a 0.78~$M_\odot$ model at three different $M_{\rm acc}$ of 0.84 (dashed line), 3.05 (dotted line), and $3.47\times 10^{-4} M_\odot$ (solid line) with $T_{\rm c} = 5\times 10^6$~K. The core-envelope boundary shifts inwards as $M_{\rm acc}$ increases. This systemic evolution gives rise to the relative increase in the envelope contribution to the WKB integral compared to the core. The ratio of envelope to core integrand strength increases from approximately 1.35 to 1.55 for $M_{\rm acc} = 0.84$ and $3.47\times10^{-4}~M_\odot$, respectively. This may indicate that core modes (confined modes) should be less impacted than the envelope modes (trapped modes) and remain stable in frequency, as the deeper interior is comparatively less influenced by the changes in the accreted outer layers.

However, the envelope modes become very sensitive to the changes in $M_{\rm acc}$, hence experiencing more pronounced shifts in frequency. Furthermore, the prograde mode orders $n_{\rm g} = 5-6$ and the retrograde $n_{\rm g} = 6-7$ exhibit ``avoided mode crossing" features in Figure \ref{fig:modesvsmacc0.78M}. The sensitivity of envelope mode could potentially cause the phenomenon of avoided mode crossing \citep{Christensen-Dalsgaard_2010}, as the stability of core modes with $M_{\rm acc}$ could lead to two consecutive modes of the similar $\ell$ and $m$ to have a closest approach without crossing each other. This avoided mode crossing is due to a unique reorganization of node distribution, while an envelope mode might gain a node in the core, a core mode picks up the sensitivity to the outer layers, effectively interchanging the identities \citep{Aizenmann_1977}. Such characteristics are manifested to the non-uniformity in mode spacing and a possible mode pairing scenario. Consequently, both mode pairing and avoided crossing offers an intriguing probe to the WD interior features, particularly relevant to the boundary layer. 

Having discussed avoided mode crossings, we can now remark more fully upon a feature seen as $T_{\rm c}$ was varied in section \ref{ssec:gmodetcore}. This feature differs from a traditional ``avoided" mode crossing, resembling more of a ``super" mode crossing. As $T_{\rm c}$ increases in Figure \ref{fig:Radialeigenfunctions} one of the nodes in the $n_{\rm g}=5$ mode moves from the boundary region into the envelope, and one of the nodes in the $n_{\rm g}=3$ mode move from the core into the boundary region. This is likely so strong partially because of the change in boundary condition due to the presence and then absence of the solid core.

Figure \ref{fig:PdotPforMaccc0.78M} demonstrates the period derivative ($|\dot{P}|$), calculated between three and six months following a dwarf nova outburst, as a function of average observed mode periods for the dipole triplets of radial orders $n_{\rm g} = 1-15$. The WD models shown correspond to accreted layer masses of $M_{\rm } = 1.63,2.06, 2.48, 2.81, 3.05$, and $3.47\times10^{-4}M_\odot$ for a 0.78~$M_\odot$ with a fixed $T_{\rm c} = 5~\times~10^6$~K. All models have the same final $T_{\rm eff} = 14000$~K. With respect to a shortest driven  period of 192 s in the star's frame, the filled and open symbols indicate whether the modes are likely to be observed or not. The retrograde modes reveal larger $|\dot{P}|$ values compared to their prograde and zonal counterparts. While the lower-order $g$ modes show minimum sensitivity in $|\dot{P}|$ to changes in $M_{\rm acc}$, the higher-order modes display discernible variations. Particularly, the sudden decrease in $|\dot{P}|$ for the prograde mode orders $n_{\rm g}= 5-6$ can be attributed to the previously discussed ``avoided mode crossing" behavior, as illustrated in Figure \ref{fig:modesvsmacc0.78M}. A similar feature is also observed for the retrograde mode orders $n_{\rm g}=6-7$, as seen in the top panel of Figure \ref{fig:PdotPforMaccc0.78M}. That is, one of these modes is the ``trapped" envelope mode, and will thus have a high $\dot{P}$, and other is not. Within the framework of the known mode driving mechanism for accreting WDs \citep{Arras_Townsley_Bildsten_2006}, and considering the shortest driven period adopted in this work, we find that nearly all retrograde $g$ modes up to radial order 12 are expected to be driven and thus observable, whereas only the lowest order $n_{\rm g}=1$ prograde mode is likely to be excited, as illustrated in Figure \ref{fig:PdotPforMaccc0.78M}. This suggests that retrograde $g$ modes are the most likely candidates for detection in accreting WDs and similar systems \citep{Kumar_Townsley_2026}.

\begin{figure}[h]
  \includegraphics[width=1.\columnwidth]{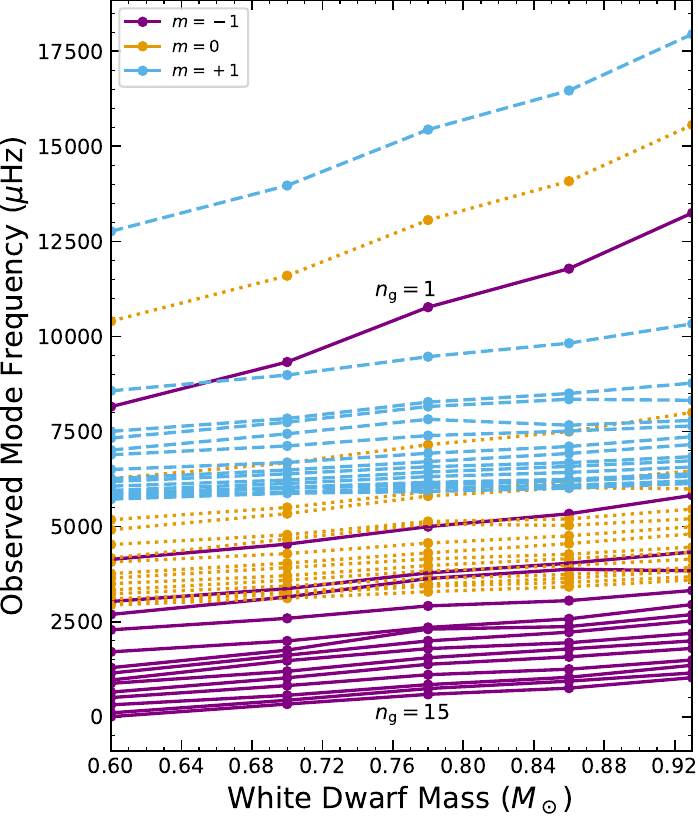}
   \caption{Pulsation frequencies of the first 15 dipole $g$-modes in terms of the stellar mass for the models with $T_{\rm c}=6\times10^6$~K at $M_{\rm acc} = 2/3~M_{\rm ign}$. The lower-mass models have a thicker envelope size. The mode frequency increases with increasing $M_{\rm WD}$.}
   \label{fig:gmodemass} 
\end{figure}

\subsection{Gravity Modes vs. White Dwarf Masses} \label{ssec:gmodewdmass}
In this section, we present our $g$ mode calculations for various WD masses. Specifically, we examine five different WD mass models with $M = 0.60,0.70,0.78,0.86$, and $0.93~M_\odot$. For each model, we compute the dipole ($\ell=1)$ $g$-mode structure during the long-term accretion phase when $M_{\rm acc} = 2/3~M_{\rm ign}$ (as shown in Table \ref{table:table_all_parameters})at $T_{\rm c} = 6\times 10^6$~K. Element diffusion is fully taken into account in our computations. The accretion rate is chosen so that the $T_{\rm eff}$ is the same for all models at the end of the long-term accretion, which is 14000~K. The effect of stellar mass on pulsation frequencies is displayed in Figure \ref{fig:gmodemass} for the first 15 radial orders. The dashed blue, dotted orange, and solid purple lines represent prograde, zonal, and retrograde modes, respectively. Overall, the mode frequencies increase (i.e., mode periods decrease) with increasing $M_{\rm WD}$. This trend is expected, as the higher WD mass possesses higher gravity and, in turn, has a larger Brunt-V\"ais\"al\"a frequency compared to the lower masses, hence exhibiting higher pulsation frequencies. Notably, even though the most massive model has the smallest accreted envelope, the frequency spacing between the consecutive modes is the largest. 
\begin{figure}[h]
  \includegraphics[width=1.\linewidth]{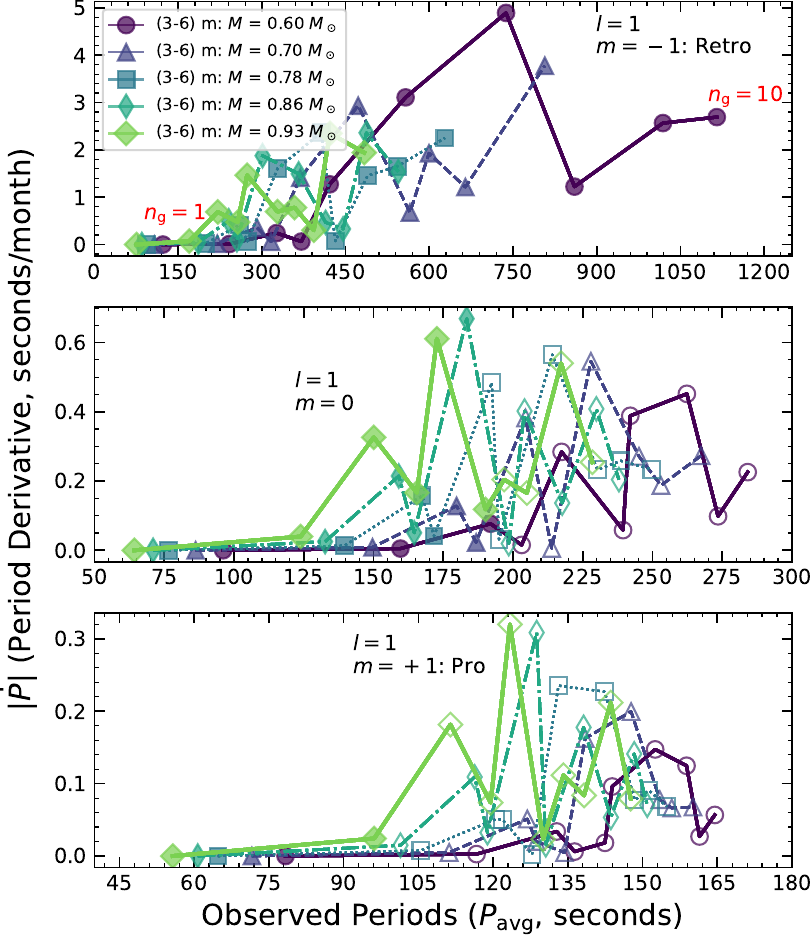}
   \caption{Similar to Figure \ref{fig:PdotPforMaccc0.78M} but for 0.60~$M_\odot$, 0.70~$M_\odot$, 0.78~$M_\odot$, 0.86~$M_\odot$, and 0.93~$M_\odot$ WD models, in the observer's frame of reference. The top, middle, and bottom panels describe for $m=-1$, $m=0$, and $m=1$ of $\ell=1$ triplets. }
   \label{fig:PdotPfordiffmass} 
\end{figure}

Figure \ref{fig:PdotPfordiffmass} explores $|\dot{P}|$ as a function of observed mode period for various WD mass models following a dwarf-novae event, similar to Fig. \ref{fig:PdotPforMaccc0.78M}. Lower-mass models possess longer retrograde mode periods and also show greater $|\dot{P}|$. In particular, for the 0.60~$M_\odot$ model, mode orders $n_{\rm g} = 4$ to 7 show a continuous increase to $|\dot{P}|$, peaking at $n_{\rm g}=7$ in both retrograde and prograde modes. This behavior, however, is not evident in the higher mass models. The lower-order modes show the smallest variations to $|\dot{P}|$ across all the WD models. 

The zonal modes ($m=0$) show the role of the chosen driving threshold more clearly, since they are traveling neither with nor against the rotation. The most massive (0.93~$M_\odot$) WD model exhibits the large number of driven zonal modes, with six modes excited. However, the lowest WD mass model shows only three zonal modes are excited. The higher mass model has both a shorter frequency $n_{\rm g}=1$ mode and a smaller mode spacing so a fixed driving threshold period will give more driven modes. Despite having no influence of rotation, the zonal modes have shorter observed periods than the retrograde modes, hence, the lowest order retrograde modes remain favorable to detections.

\begin{figure}[ht]
  \includegraphics[width=1.\linewidth]{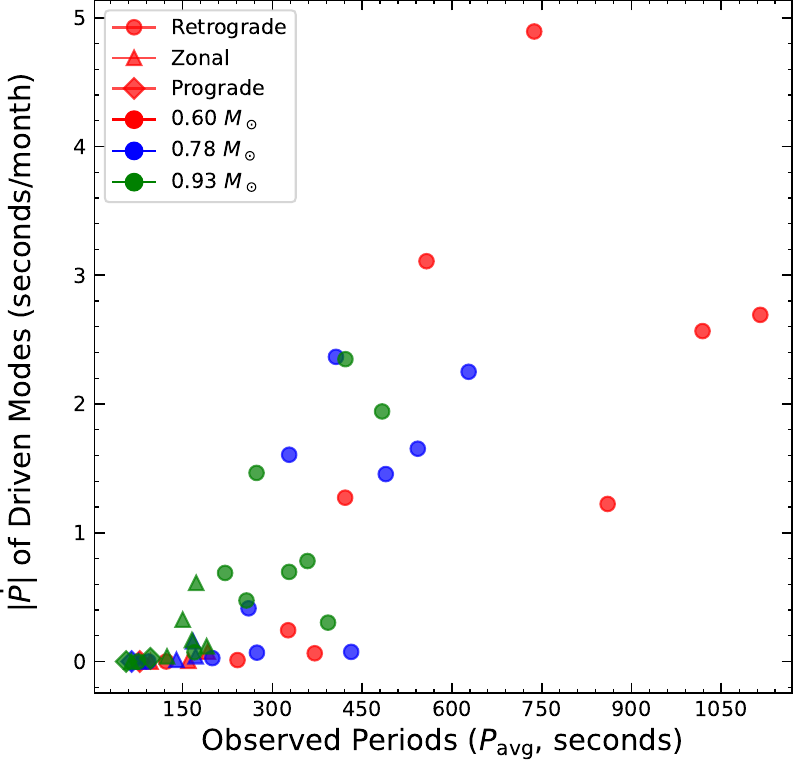}
   \caption{$|\dot{P}|$ as a function of observed mode periods for the driven modes shown in Figure \ref{fig:PdotPfordiffmass} in WD models with $M_{\rm WD} = 0.60,0.78$, and $0.93~M_\odot$, indicated by red, blue, and green colors, respectively. Retrograde, zonal, and prograde modes are shown with circle, triangle, and diamond symbols, respectively.  The minimum driving period of 192 seconds is adopted in star's frame of reference.}
   \label{fig:Pdotdrivenmodes} 
\end{figure}

\autoref{fig:Pdotdrivenmodes} presents the $|\dot{P}|$ values of the driven modes (up to $n_{\rm g}=10$) as a function of their observed periods for three different WD models with $M_{\rm WD} = 0.60, ~0.70$, and $0.93~M_\odot$, extracted from Figure \ref{fig:PdotPfordiffmass}. Among all driven modes, retrograde modes consistently evidenced the largest $|\dot{P}|$. This is because the retrograde modes have longer periods hence, a given fractional change in periods leads to a larger $|\dot P|$. This makes them more susceptible to observations than the prograde or zonal modes of the same order. Furthermore, figure \ref{fig:Pdotdrivenmodes} also shows that only the lowest order prograde modes are effectively driven and exhibit the smallest $|\dot{P}|$. 

\section{Discussion and Conclusions} \label{sec:discussions}
The seismology of WDs in CVs is relatively less well understood than that of isolated WDs due to several challenges, including their being subject to frequent outbursts, very faint observational signatures, and the intrinsic difficulty in modeling their exact evolution \citep{Schreiber_Zotorovic_Wijnen_2016, Shen_2022}. In this work, we have presented the first comprehensive seismological analysis of the accreting WDs  using $g$ modes. We use a forward modeling approach rather than trying to fit the observed modes in a particular star. To this end, we choose representative values of the parameters characterizing the WD. Specifically, we investigate three main modeling components that shape these systems: the core temperature, the mass of the accreted H-rich layer ($M_{\rm acc}$), and the WD mass. We envisage that $g$ modes are useful for probing these properties in rapidly rotating accreting WDs. While previous studies proposed that Rossby modes ($r$ modes) may be more favorable in such systems \citep{Saio_2019}, recent work \citep{Kumar_Townsley_2026} demonstrated that $g$ modes are more likely to dominate pulsations in accreting WDs. 

We compute $g$-mode frequencies for low-order modes during quiescence as well as their variation following a dwarf nova outburst for the underlying WD models summarized in Table \ref{table:table_all_parameters}. The period spacings change in distinctive ways for each variation in the parameter of the WD. The variation of $T_{\rm c}$ seems to show the most change in which order mode shows minimum period spacing. 
The variations of $M_{\rm acc}$ have the largest effect on the $n_g=1$ mode and lead to modest changes in the order of the mode with minimum mode spacing. A change in mass leads mostly to a general increase in all mode frequencies, but low-order modes are the most affected. Furthermore, the presence or absence of a solid core doesn't seem to have a strong effect, based either on variation of $T_{\rm c}$ or mass, but it shows some potential significance. While the higher-order $g$ modes generally follow the asymptotic relation, $\Pi\propto(\int_{r1}^{r2} |N|dr/r)^{-1}$ \citep{Tassoul_1990}, deviations from this relation are still apparent. This indicates that the crystallization and structural changes modify the finite departure from the ideal asymptotic relation. This is explained by using the WKB integrand shown in Figure \ref{fig:wkbcoretemp}, as the envelope to core integrand decreases as $T_{\rm c}$ increases, leading to systematic changes in $N/r$. The fact that some radial orders frequencies are strongly affected by the changes in the extent of the crystallized core while others are not, indicates that we may be able to possibly unwind the impact of crystallization with the help of various surface layer masses. 

We note that more recent studies have proposed that the WD crystallization may generate magnetic fields \citep{Isern2017}, potentially transforming CV to magnetic systems \citep{Schreiber2021}. While current models suggest that thermohaline mixing associated with crystallization is too weak to sustain a strong magnetic dynamo \citep{Montgomery2024}, convective motions may still be able to dredge-up previously buried magnetic fields \citep{Blatman2024a, Blatman2024b} and the presence of such magnetic fields could, in turn, suppress the lower-frequency $g$ modes \citep{Rui2025}. However, in this work, we do not consider any magnetic effects, and our results therefore apply most directly to non-magnetic CVs. A self-consistent treatment of crystallization-driven magnetism and its impact on pulsation modes should be investigated in detail for accreting systems in future work.

The general and qualitative trend of a particular mode after the dwarf nova remains similar to what is presented in \citet{Kumar_Townsley_2023}, relaxing to its pre-outburst value in a few months time. However, the WD models presented here have significant improvements from our previous work.

Identifying the observed modes with the theoretically predicted ones has been challenging. We generally rely on the spacing between consecutive modes in order to match the observations. To better constrain the mode identification, we have presented a new method to identify mode orders using $|\dot P|$ verses $P_{\rm avg}$ diagram. Using this, we encapsulate the individual mode variations following the strong accretion outburst allowing us to focus on behavior of the single mode. We see that the placement of modes in the $|\dot{P}|-P$ plane is distinctive to certain parameters. The modes that show the most promise from this method are the retrograde modes, because their long periods give a larger $|\dot P|$ for a given fractional rate of change in the period. These modes are also more likely to be driven in a typical driving scenario. This suggests that we might be able to identify modes without having a complete mode sequence, but more study is required. This aims to provide a new foundation for interpreting mode structures and thereby better constraining the interior of the accreting WDs. 

We also note that the low-frequency $g$ modes in WDs can exhibit stochastic variations in their frequencies \citep{Montgomery2010, Hermes2017}, and this variability may introduce additional scatter in the observed mode periods. As mentioned by \cite{Hermes2017} for the DAV case, such variability is mostly seen with periods larger than 800 second. Those observations pertain to pure hydrogen atmosphere WDs, whereas the system of interest here involve accreting white dwarfs with mixed atmosphere, which might further complicate things. Despite this, we believe that the overall trends underlying our proposed observables should remain largely unchanged.

The relevant input files and datasets are available
online at doi: \href{https://doi.org/10.5281/zenodo.18307075}{https://doi.org/10.5281/zenodo.18307075}

\section*{Acknowledgements}
This manuscript has benefited from insightful comments by the anonymous referee. We thank Ken Shen, Alan Calder, Sam Boos, and Nethra Rajavel for the useful discussions. We also would like to thank Spencer Caldwell and Broxton Miles for their earlier work on setting up the nova simulation. This work was supported under programs HST-GO-15072, HST-GO-16069, and HST-AR-16638 through the Space Telescope Science Institute, which is operated by the Association of Universities for Research in Astronomy, Inc., under NASA contract NAS5-26555. Support for these programs was provided through a grant from the STScI under NASA contract NAS5-26555.
\vspace{5pt}

\textit{Software}: \texttt{MESA} (\citealt{Paxton_2011, Paxton_2013, Paxton_2015, Paxton_2018, Paxton_2019}, \href{https://docs.mesastar.org/en/latest/}{https://docs.mesastar.org/en/latest/}),\

\texttt{GYRE} (\citealt{Townsend_2013, Townsend_2018}, \href{https://gyre.readthedocs.io/en/stable/}{https://gyre.readthedocs.io/en/stable/}),\

\texttt{Matplotlib} (\citealt{Hunter_2007}, \href{https://matplotlib.org/}{https://matplotlib.org/})

\section*{ORCID iDs}
\noindent
Praphull Kumar: \orcidlink{https://orcid.org/0000-0002-8791-3704}\href{https://orcid.org/0000-0002-8791-3704}{0000-0002-8791-3704}\\
Dean M. Townsley: \orcidlink{https://orcid.org/0000-0002-9538-5948}\href{https://orcid.org/0000-0002-9538-5948}{0000-0002-9538-5948}

\nopagebreak
\bibliography{references}{}

\bibliographystyle{mnras}

\end{document}